\documentstyle[preprint,aps,a4,12pt,axodraw]{revtex}
\begin{document}
\draft
\hfill\vbox{\baselineskip14pt
            \hbox{\bf KEK-TH-507}
            \hbox{\bf YUMS-97-002}
            \hbox{\bf SNUTP-97-009}
            \hbox{(\today)}}
\baselineskip18pt
\begin{center}
{\bf Is CDF large $p_{_T}$  anomaly from virtual 
SUSY threshold effects?}\footnote{Talk is given at the YITP International Workshop -- 
`Recent Developments in QCD and Hadron Physics', on Dec. 16--18, 
at Kyoto Japan. Proceedings will be published, edited by Y. Koike.}
\end{center} 
\begin{center}
\large C.S. Kim$^{1,2}$
\end{center}
\begin{center}
1: {\it Department of Physics, Yonsei University, Seoul 120-749, Korea}\\
2: {\it Theory Group, KEK, Tsukuba, Ibaraki 305, Japan }\\
\end{center}
\begin{center} 
\large Abstract
\end{center}
\begin{center}
\begin{minipage}{14.5cm}
\baselineskip=15pt
\noindent

Talk is given at the YITP International Workshop -- 
`Recent Developments in QCD and Hadron Physics'.
Recent CDF data  of the inclusive jet cross section shows 
anomalous deviation around large
transverse momentum $p_{_T}(j)\approx 200 \sim 350 $ GeV. 
Is it possible to interpret the anomaly in terms of
virtual SUSY effects. The answer is `NO', because
we find that the virtual SUSY loop  interference effects 
are too small to explain the CDF data.    

\end{minipage}
\end{center}
\baselineskip=18pt
\normalsize

\section{Introduction}

The CDF [1] and D0 [2] collaborations  at Tevatron 
Collider have recently reported data for the inclusive jet cross section 
in $p\bar{p}$ collisions at $\sqrt{s} = 1800$ GeV. 
Let us recapitulate some particulars of this data. 
\begin{itemize}
\item[(a)] The CDF measurement is based on a data sample of 19.5 pb$^{-1}$
collected in 1992-93 with the CDF detector at the Tevatron collider.
Jets were reconstructed using a cone algorithm.
\item[(b)] The measurements have been reported over a wide range 
of transverse energy, 15 GeV $\leq p_{_T} \leq$ 440 GeV, 
and around the central pseudorapidity region 
${\rm 0.1 \leq |\eta| \leq 0.7}$. The smallest distance 
probed is on the order of $10^{-19}$ m.
\item[(c)] After accounting for uncertainties the corrected
experimental cross section, when compared to the 
Next-to-leading [NLO] QCD predictions using MRSD0$^\prime$ parton 
distribution function [PDF's], is significantly higher than
the NLO prediction for $p_{_T} > 200$ GeV. 
For $p_{_T}$ below 200 GeV the agreement between the CDF and the NLO QCD
cross section is excellent, while the D0 results are 
higher than the NLO PQCD  
predictions within the statistical uncertainties.
\end{itemize}  

There are basically two logical possibilities
for explanations of the CDF data on inclusive jet production cross sections:
(i) the parton distribution functions determined at low $Q^2$ region may not
be accurate enough to be applicable to the high $p_{_T}$ region with $p_{_T} > 
200$ GeV, or (ii) there are some new physics around the electroweak scale.
We give a brief review of  these two possibilities in the following.

The most interesting possibility is that jet measurements at hadron
colliders may be sensitive to quantum corrections due to
virtual SUSY particles [3,4,5]. The
purpose of this note is to concentrate on this scenario.
The layout of this paper is as follows. In next section we discuss
our calculation of the SUSY virtual threshold effects. The final
section contains discussions and conclusions of our numerical
results.

\section{Virtual SUSY Threshold Effects}

It is well-known that SUSY particles [gluinos, squarks] decrease
or slow down the rate of fall of $\alpha_s(\mu)$ for large scale $\mu$.
``Large'' means far above the threshold. At the one loop
level the evolution equation for $\alpha_s(\mu)$ can be written as
\begin{equation}
\frac{d}{d\ln \mu}\alpha_{s}(\mu)=-\frac{b_{3}}{2\pi}
\alpha_{s}^{2}(\mu)~.
\end{equation}
In the SM, $b_3$ is given by
\begin{equation}
b_3=11-\frac{2}{3}n_f~,
\end{equation}
whereas in MSSM model one has
\begin{equation}
b_3=11-\frac{2}{3}n_f-2-\frac{1}{3}\tilde{n}_f~,
\end{equation} 
where $n_f$ [$\tilde{n}_f$] is the number of quark [squark] flavors that
are active. The contribution `$-2$' is the gluino contribution [it is assumed
that the gluino is active in this case]. 
In Ref. [4] the issue of the
effect of high-mass thresholds due to gluinos, squarks and other
new heavy quarks on the evolution of $\alpha_s$ was considered.
The corrections to $\alpha_s$ were found to be appreciable, this
in turn means a significant increase in the transverse momentum
dependence of jet production at the Tevatron. However, as noted
in Ref. [5], these  authors [4] do not include
the effect of $q \tilde{q} \tilde{g}$ Yukawa interactions 
and hence one cannot take their
estimates for the superpartners of ordinary matter as final. 
In Ref. [5] the effect of Yukawa couplings was included.
The results in [5] indicate that the CDF data cannot
be explained by a mass threshold effect in the MSSM, as the
calculated result is not only small but of the wrong sign.
In a similar but more detailed analysis, the authors of
Ref. [3], working in the context of MSSM, consider
the virtual one-loop corrections to the parton-level subprocesses
$q\bar{q}\rightarrow q\bar{q}$, 
$q\bar{q}\rightarrow q^\prime \bar q^\prime$,
$qq^\prime\rightarrow qq^\prime$, 
$q\bar{q}\rightarrow gg$ and $qg\rightarrow qg$, 
which are expected to dominate the large $p_{_T}$ cross sections at
the Tevatron energy.

The purpose of this talk is to give our results
of incorporating the one-loop radiative corrections into
the running of $\alpha_s$, the dressing-up of the parton
distribution functions, and finally convoluting the relevant
subprocess cross sections with the SUSY dressed-up PDF's. 
We note that  one 
should take into account sparticles effects on the parton
structure functions at energies sufficiently far above
the threshold and can ignore this effect in the threshold
region. In the hadron colliders, like the 
Tevatron, what is measured is $p\bar{p}$ cross section, 
and not the individual subprocesses cross sections.
So in order to determine the effect of
subprocesses on the $p_{_T}$ cross section one must perform
a convolution of the cross section of each subprocess
with the corresponding PDF's. We find that in the process 
of convolution with the PDF's, the ``dips and peaks'' in the various 
subprocesses [3] are much reduced.

\section{Discussions and Conclusions}

For our numerical calculation, we implement various lower bounds 
on squarks and gluinos depending on parameters in the MSSM.  For example, 
D0 group [6] searched for the events with large missing $p_{_T}$ 
with three or more jets, observed no such events above the level 
expected in the SM.
This puts some limits on the squark and gluino masses assuming 
the short-lived gluinos:
\begin{eqnarray}
m_{\tilde{g}} & > &  144 ~{\rm GeV~~~for~~~} m_{\tilde{q}} = \infty~,
\\
{\rm or}~~~~m_{\tilde{g}} & = & m_{\tilde{q}} > 212~{\rm GeV}~. 
\label{eq8}
\end{eqnarray}
CDF group is currently analyzing their data, with their
preliminary data being similar to the D0 results  with slight increase in
sparticles' mass bound. 
In the subsequent numerical analysis, we choose three sets of 
$(m_{\tilde{g}}, m_{\tilde{q}})$, which we shall refer to as 
Case I, II and III, respectively,
\begin{equation}
(m_{\tilde{g}}, m_{\tilde{q}}) =  (220 ~ {\rm GeV}, 220~ {\rm GeV}),~~
(150~ {\rm GeV}, \infty),~~ (150 ~ {\rm GeV}, 150~ {\rm GeV})~.
\label{eq9}
\end{equation}
The Case III is only of academic interest if the limit given in 
Eq. (\ref{eq8}) is valid in reality.
 
Fig.~1 exhibits our results for $[{\rm (\frac{MSSM-SM}{SM})
\frac{d\sigma}{dp_{_{T_j}}}}]$ versus ${\rm p_{_{T_{j}}}}$ for
both Cases I and III [Case III refers the MSSM with the five squarks 
and gluinos being degenerate with a common mass of 150 GeV, {\it i.e.} 
the last set in Eq. (\ref{eq9})]. 
A rapidity cut of 0.7 is used [$|\eta|<0.7$].
A maximum difference of $-1\%$ from the SM is found for the both 
Cases. The question that naturally arises is, that why the  
``dips and peaks'' [which are at the level of 5\%--6\% [3]]
at the parton level, are reduced. 
There are two factors contributing to the reduction in the
``dips and peaks'' in the process of going from the parton to the 
process level. Part of this reduction comes from the process of
convolution of the various subprocesses with PDF's [as already 
remarked to in the previous section].  
The other piece of this reduction in ``dips and peaks''
by going from subprocess to process level is due to t-channel
``dilution'' effect. This can be shown by simply not including
the t-channel subprocesses' contribution. When this is done [see
Fig.~2] reduction in the ``peaks and dips ''is not so large.

In Fig.~2 we exhibit our results for $[{\rm (\frac{MSSM-SM}{SM})
\frac{d\sigma}{dp_{_{T_j}}}}]$ versus ${\rm p_{_{T_{j}}}}$ for
both Cases I and III, but this time only including the subprocesses
$q\bar{q}\longrightarrow q'\bar{q}'$ and $q\bar{q} \longrightarrow gg$.
These two are the subprocesses with the prominent threshold structures from
vacuum polarization interferences through $s$-channel exchange diagrams. 
A rapidity cut of 0.7 is used [$|\eta|<0.7$].
The deviation from the SM for Case I varies between 2\%  to $-4\%$,
whereas as in Case II the variation is similar size to Case I at 
different energy scales related to the SUSY threshold effects. 

In summary, in the MSSM the $p_{_T}$ distributions does not
differ very much from those of the SM except for the possible 
threshold effects ($\sim 1\%$) through loop corrections.  
In actual experiments, the jet resolution
will in general smear out any narrow resonance structures (which may be the
case for the long-lived gluinos), leading to broad resonance structure,
and therefore it looks impossible to detect SUSY particles through this kind
of indirect virtual threshold effect.  
It has been reported [7] that
the apparent discrepancy between CDF data and theory may be 
explained by the uncertainties resulting from the non-perturbative
parton distribution, in particular in the gluon distribution.
Our main conclusion seems to be on the right track in view
of this  global analysis of parton distribution function.
For more details on this talk, please look Ref. [8].

\section*{Acknowledgements}

The work of CSK was supported 
in part by KOSEF, 
Project No. 951-0207-008-2, in part by CTP of SNU, 
in part by the BSRI Program, Project No. BSRI-97-2425, and
in part by the COE Fellowship from Japanese Ministry of 
Education, Science and Culture.


\section*{references}

\begin{itemize}

\item[[1]] CDF Collaboration: F. Abe {\it et al.}, 
Phys. Rev. Lett. {\bf 77} (1996) 438.
\item[[2]] D0 Collaboration: G. Blazey, talk given at Recontre de
Moriond, (March 1996).
\item[[3]] J.~Ellis and D.~Ross,~Phys.~Lett.~{\bf B383}~(1996)~187.
\item[[4]] V. Barger {\it et al.}, Phys. Lett. {\bf B383} (1996) 178.
\item[[5]] P.~Kraus and F. Wilczek,~Phys.~Lett.~{\bf B382}~(1996)~262.
\item[[6]] D0 Collaboration: S. Abachi {\it et al.}, Phys. Rev. Lett.
{\bf 75} (1995) 618.
\item[[7]] H.L.~Lai {\it et al.}, hep-ph/9606399.
\item[[8]] C.S. Kim and S. Alam, hep-ph/9610503.
\end{itemize}

\section*{Figures}

\begin{itemize}

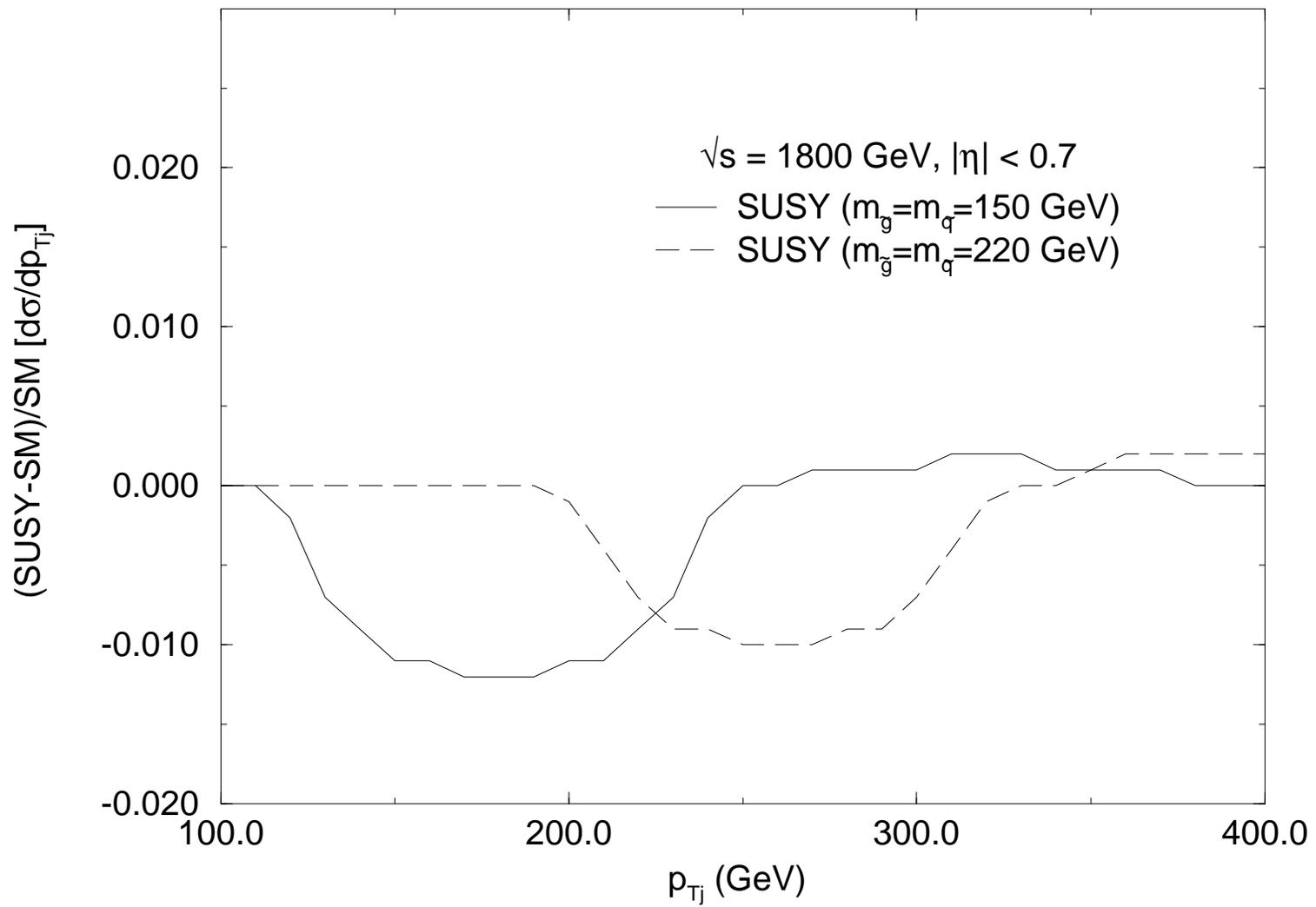
\item[{\bf Fig.~1}] Deviation from the SM due to SUSY
contribution to $\frac{d\sigma}{dp_{_{T_j}}}$ versus $p_{_{T_j}}$.
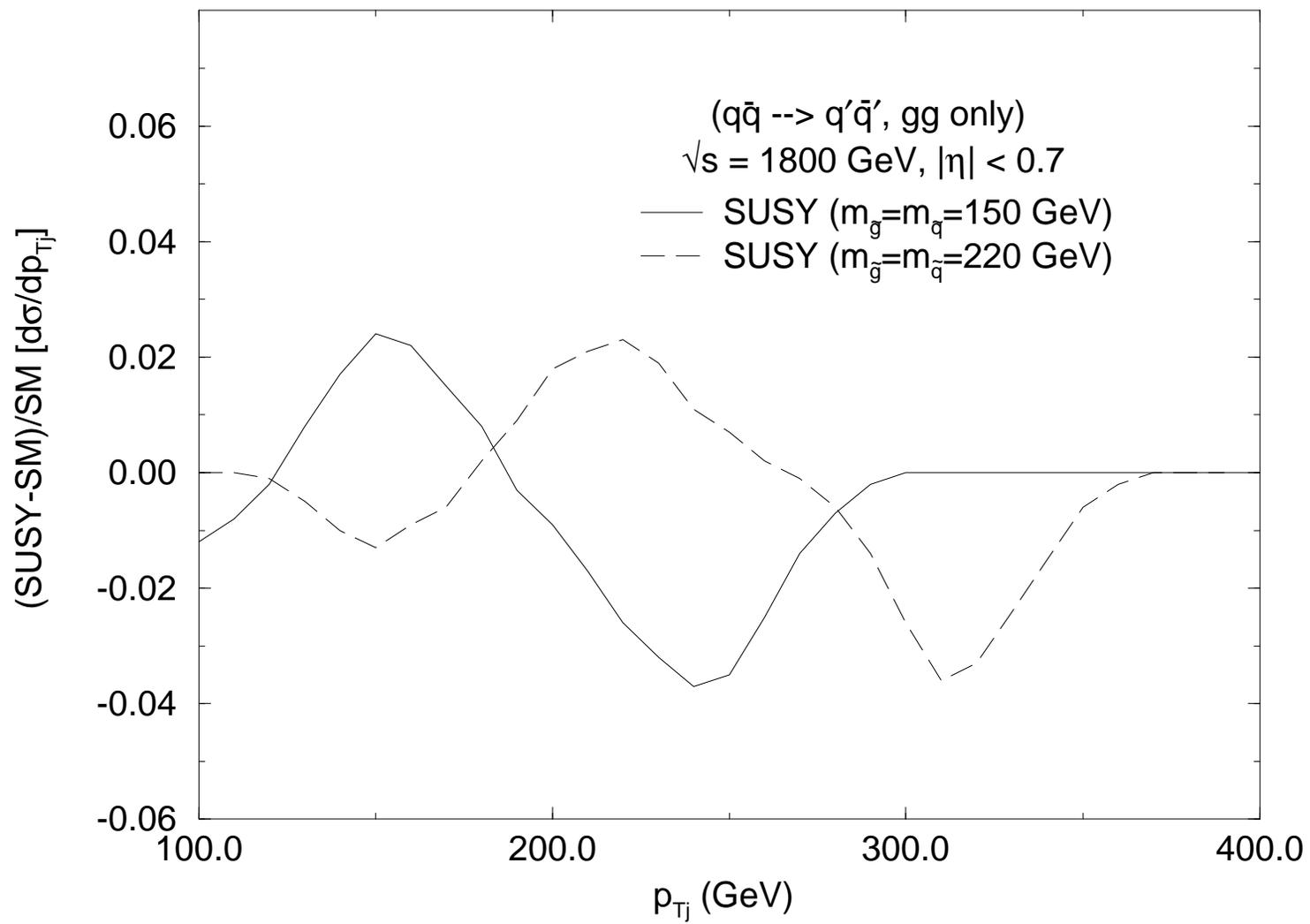
\item[{\bf Fig.~2}] Deviation from the SM due to SUSY
contribution to $\frac{d\sigma}{dp_{_{T_j}}}$ versus $p_{_{T_j}}$, when only 
the subprocesses $q\bar{q}\rightarrow q^{\prime}\bar{q}^{\prime}$,
and $q\bar{q}\rightarrow g g $ are included.
  
\end{itemize}

\end{document}